# Algorithmic Analysis of Invisible Video Watermarking using LSB Encoding over a Client-Server Framework

Poorna Banerjee Dasgupta

*M.Tech Computer Science & Engineering, Nirma Institute of Technology*
*Ahmedabad, Gujarat, India*

***Abstract*** — Video watermarking is extensively used in many media-oriented applications for embedding watermarks, i.e. hidden digital data, in a video sequence to protect the video from illegal copying and to identify manipulations made in the video. In case of an invisible watermark, the human eye can't perceive any difference in the video, but a watermark extraction application can read the watermark and obtain the embedded information. Although numerous methodologies exist for embedding watermarks, many of them have shortcomings with respect to performance efficiency, especially over a distributed network. This paper proposes and analyses a 2-bit Least Significant Bit (LSB) parallelized algorithmic approach for achieving performance efficiency to watermark and distribute videos over a client-server framework.

***Keywords*** — *Video Watermarking, LSB Coding, Media Copy Control.*

## I. INTRODUCTION

We live today in a world where sharing and distribution of digital media such as songs, photos and videos have become very popular. With this rapid increase in sharing and distribution, comes the problem of digital media copying and piracy. As mentioned earlier, video watermarking refers to the process of embedding hidden digital data in a video. Ideally in case of an invisible watermark, a user viewing the video cannot perceive a difference between the original, unmarked video and the marked video, but a watermark extraction application can read the watermark and obtain the embedded information. The extracted information can then be used to determine the authenticity of a video as well as help in differentiating between an original and a copied video. Some common applications of video watermarking include [1],[6]:

- *Copyright Protection*: For the protection of intellectual property, the video data owner can embed a watermark representing copyright information in video data. This watermark can help prove ownership in a legal court when someone has infringed on the owner's copyrights. For instance, embedding the original video clip by noninvertible video watermarking algorithms during the verification procedure helps to prevent the multiple ownership problems in some cases.

- *Video Authentication*: Popular video editing software's available today permit users to easily tamper with video content. Authentication techniques are consequently needed in order to ensure the authenticity of the content. One solution is the use of digital watermarks. Timestamp, camera ID and frame serial number are used as a watermark and embedded into every single frame of the video stream.

- *Video fingerprinting*: To trace the source of illegal copies, a fingerprinting technique can be used. In this application, the video data owner can embed different watermarks in the copies of the data that are supplied to different customers. Fingerprinting can be compared to embedding a serial number in the data that is related to the customer's identity. It enables the intellectual property owner to identify customers who have broken their license agreement by supplying the video data to third parties.

- *Copy control*: The information stored in a watermark can be used to directly control digital recording devices for copy protection purposes. In this case, the watermark represents a copy-prohibit bit and watermark detectors in the digital-media recorder determine whether the video data offered to the recorder may be stored or not.

This paper presents a simple yet very efficient algorithmic approach for invisibly watermarking videos using the concept of 2-bit Least Significant Bit (LSB) encoding/decoding over a client-server framework. The following sections of this paper describe in detail how the structure of a video file can be deconstructed to isolate particular video frames for watermarking and how the encoding/decoding modules of the watermarking algorithm work using Base64 coding.

## II. IDENTIFYING I-FRAMES FROM A VIDEO SEQUENCE

Video files, such as those coded by the MPEG and H.262/H.263 standards comprise of information headers followed by a sequence of image frames [2],[3]. For the purposes of clarity and demonstration, all video codings hereby mentioned will refer to the MPEG-1 standard. The video header bit-stream consists of a hierarchical structure with seven layers [4]:

- Video Sequence
- Sequence Header
- Group of Pictures (GOP)
- Picture
- Slice
- Macroblock (MB)
- Block

A video is identified by a sequence. Data for each sequence consists of a sequence header followed by data for group of pictures (GOPs). The sequence header comprises of various fields such as *aspect ratio, frame rate, vertical and horizontal sizes* of frames and a *32-bit sequence-header code* which has the value 000001B3 to indicate the beginning of the sequence header.

Data for each Group of Pictures consists of a GOP header followed by Picture data. The GOP header comprises of various fields such as *time-code, closed-gop, broken-link and a 32-bit group-start-code* having value 000001B8 which indicates the beginning of a Group of Pictures.

The Picture layer consists of a picture-header followed by consecutive picture frames. These picture frames are classified as follows:

- *I-frame (intra coded picture)*: a picture frame that is coded independently of all other frames. Each GOP begins (in decoding order) with this type of frame.
- *P frame (predictive coded frame)*: contains motion-compensated difference information relative to previously decoded frames. A P-frame may refer to only one other preceding picture frame.
- *B frame (bi-predictive coded frame)*: this frame also contains motion-compensated difference information relative to previously decoded pictures but as the name indicates, such a frame can refer to two other picture frames – one preceding and the other one succeeding the B-frame.
- *D-frame (DC direct coded picture)*: serves as a fast-access representation of a picture for loss robustness or fast-forwarding.

Continuing the above mentioned hierarchy, the Picture layer consists of a picture header followed by slice data. Among other fields in the picture-header, the most important ones are the *picture-start-code* (with a 32-bit value of 00000100 to indicate the beginning of a picture frame) and the *picture-coding type*, which is a 3-bit value indicating the frame type. Figure 1 shows the Picture layer header format [4].

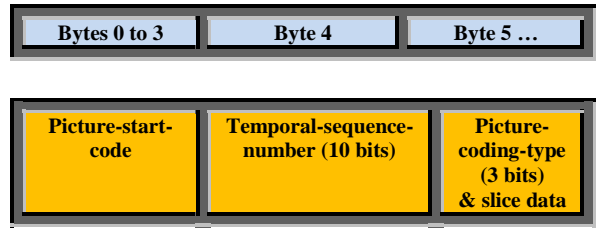

Fig. 1 Picture Layer Header Format

Based on the value of the *picture-coding-type*, one can differentiate between an I-frame, P-frame, B-frame and a D-frame. The value of *picture-coding-type* is 001 for an I-frame, 010 for a P-frame, 011for a B-frame and 100 for a D-frame. Since the first frame in every GOP starts with an I-frame and because I-frames contain important video information, I-frames are ideally suited for being watermarked.

In the algorithmic approach proposed in this paper, consecutive I-frames have been identified and selected for being watermarked and a checksum of that watermark has been added to the I-frames for further authentication and security. The following sections describe how the actual encoding/decoding of the watermark has been done on the I-frames as proposed.

## III. ENCODING/DECODING THE WATERMARK IN VIDEO FRAMES

As explained in the previous section, encoding and decoding of the watermark has been done on consecutive I-frames. The watermark itself has been chosen to be a document file which contains text and logos to describe the authentication and ownership rights of the video. The proposed watermarking algorithm works in three phases:

1. *Encoding the watermark*: A normal unmarked video serves as the input to the watermark encoder. During encoding, whenever an I-frame is detected, the RGB color pixel values of that I-frame will be stored in a matrix format. Each colored pixel requires24 bits for storage with 8 bits being allocated for each color component. Now in order to embed the watermark text file, each character from the text file is read in the form of a bytestream. Using Base64 [7] encoding, the 8 bit unsigned character value, denoted by **C,** can be converted to an equivalent 6 bit value **V**, which is then divided into 3 pairs of 2 bits each – **V1, V2, V3**. These 2-bit values are then logically exclusive-ORed (XOR) with the color components of the pixel, i.e. – V1 is XORed with R color component, V2 is XORed with G color component and V3 is XORed with B color component. This process is continued till every character in the watermark text file has been read and embedded in the I-frame and the entire procedure is repeated for other I-frames. In order to improve the robustness, a checksum value of the watermark text file is also added to each I-frame. The task of encoding is performed at the server end of the framework.

2. *Distributing the watermarked video*: After completing the encoding process, the server sends the watermarked video and Base64 encoded character matrix to a compliant client over a secure network. The client then begins decoding the watermarked video.

3. *Decoding the watermark*: After the client receives the watermarked video, the decoding process begins. Again, when an I-frame is detected, each color pixel is decomposed into its RGB components. The Base64 encoded matrix is read as a byte stream and each read group of 6 bits is then grouped into 3 pairs – V1, V2 and V3. These pairs are then re-XORed with the 8 bit color components, i.e. - V1 is XORed with **R'** color component, V2 is XORed with **G'** color component and V3 is XORed with **B'** color component where R',G' and B' are the watermarked pixel color components. This operation restores back the original RGB values of each pixel and this entire process is repeated for every watermarked I-frame. Also, the Base64 encoded watermark matrix is decoded and a checksum is computed for the received and decoded watermark. This checksum is then compared with the embedded checksum value in the video. Any mismatches in the two values would indicate that the video may have been manipulated during transit.

Figure 2 shows a schematic diagram depicting the watermark encoding/decoding process.

### IV. ANALYZING ALGORITHM PERFORMANCE EFFICIENCY

The algorithmic approach for watermarking proposed in this paper distributes the tasks of encoding and decoding to the server-end and client-end respectively. To make up for the time lost in network delays during sending and receiving videos by making the encoding/decoding processes highly efficient. To do so, the algorithm has been designed to be parallely operated upon since individual pixel watermarking operations are independent of each other and the data-structures involved in the algorithm are all stored and manipulated in the form of matrices. Due to such parallelization, the encoding/decoding processes can be carried out on many-core General Purpose Graphic Processing Units (GPGPUs) [5], making the performance speed-up incredibly large. In a sample test-run, for a MPEG-1 video file of size 1 MB, watermark document file of size 200 KB, and the execution platform being NVIDIA Quadro 4000 (with compute capability 2.0 and 256 cores), a speed-up of the order ~75 has been achieved when compared to the normal sequential execution of the same.

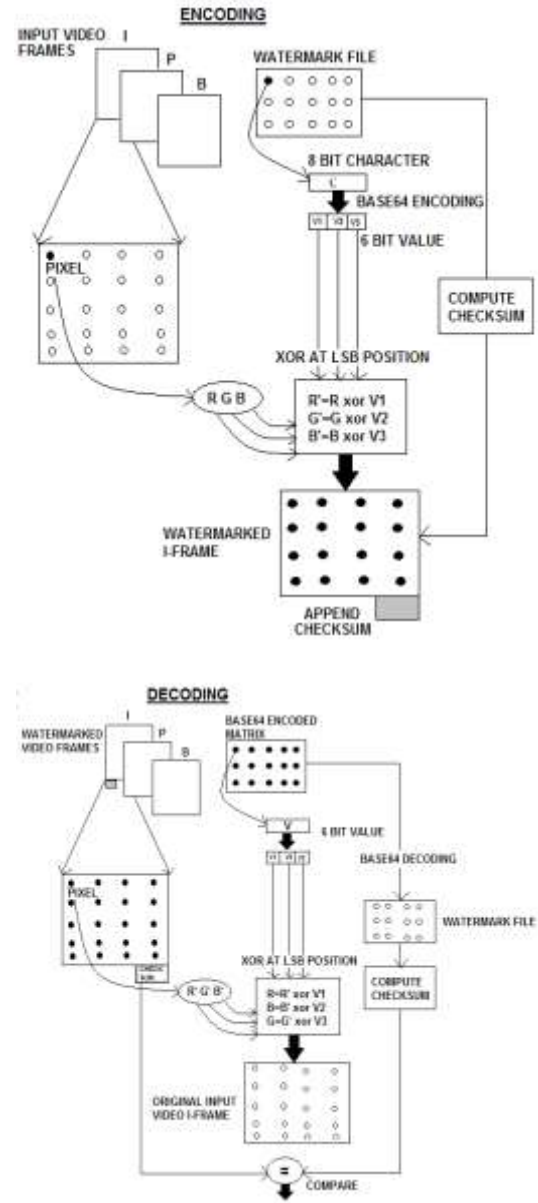

Fig. 2 Schematic of the Watermark Encoding/Decoding Process

### V. CONCLUSIONS & FUTURE SCOPE OF WORK

In today's era, sharing and distribution of digital media such as videos, photos and songs have become immensely popular. This rapid increase in multimedia sharing has caused an imminent problem – illegal multimedia distribution or piracy and copying. Watermarking such media is one of the effective solutions often employed to deal with this problem. This paper proposes and describes how invisible watermarking can be done for videos over a client-server framework using 2-bti LSB coding with checksum. To enhance the performance efficiency, the proposed algorithm and the associated data-structures have been designed in such a way so that they can be operated upon parallely with the help of many-core processors such as GPGPUs, thus greatly increasing the performance of the watermarking algorithm. This has been further confirmed by various sample test runs as described in the previous section.

As a further enhancement, the proposed algorithm can be extended to watermark videos of other coding standards such as AVI, H.263/263/264 etc. Also, algorithms in the frequency domain, such as Fourier transforms and Wavelets can be used instead for encoding/decoding the watermark.

## AUTHOR'S PROFILE

**Poorna Banerjee Dasgupta** has received her B.Tech & M.Tech Degrees in Computer Science and Engineering from Nirma Institute of Technology, Ahmedabad, India. She did her M.Tech dissertation at Space Applications Center, ISRO, Ahmedabad, India and has also worked as Assistant Professor in Computer Engineering dept. at Gandhinagar Institute of Technology, Gandhinagar, India from 2013-2014 and has published several research papers in reputed international journals. Her research interests include image processing, high performance computing, parallel processing and wireless sensor networks.